\documentclass[sigconf, nonacm]{acmart}
\newcommand\vldbdoi{XX.XX/XXX.XX}
\newcommand\vldbpages{XXX-XXX}
\newcommand\vldbvolume{14}
\newcommand\vldbissue{1}
\newcommand\vldbyear{2024}

\newcommand\vldbauthors{\authors}
\newcommand\vldbtitle{\shorttitle} 

\newcommand\vldbavailabilityurl{https://github.com/Thomas12012002/BOD}

\newcommand\vldbpagestyle{plain} 
\usepackage{amsthm}
\usepackage{algorithm}
\usepackage[noend]{algorithmic}

\begin{document}
\title{BOD: Blindly Optimal Data Discovery}

\author{Thomas Hoang}
\affiliation{
  \institution{Denison University}
  \country{USA} 
}
\email{hoang_t2@denison.edu}

\begin{abstract}
 Combining discovery and augmentation is important in the era of data usage when it comes to predicting the outcome of tasks. However, having to ask the user the utility function to discover the goal to achieve the optimal small rightful dataset is not an optimal solution. The existing solutions do not make good use of this combination, hence underutilizing the data. In this paper, we introduce a novel goal-oriented framework, called BOD: Blindly Optimal Data Discovery, that involves humans in the loop and comparing utility scores every time querying in the process without knowing the utility function. This establishes the promise of using BOD: Blindly Optimal Data Discovery for modern data science solutions. 
\end{abstract}

\maketitle

\pagestyle{\vldbpagestyle}
\begingroup\small\noindent\raggedright\textbf{PVLDB Reference Format:}\\
\vldbauthors. \vldbtitle. PVLDB, \vldbvolume(\vldbissue): \vldbpages, \vldbyear.\\
\href{https://doi.org/\vldbdoi}{doi:\vldbdoi}
\endgroup
\begingroup
\renewcommand\thefootnote{}\footnote{\noindent
This work is licensed under the Creative Commons BY-NC-ND 4.0 International License. Visit \url{https://creativecommons.org/licenses/by-nc-nd/4.0/} to view a copy of this license. For any use beyond those covered by this license, obtain permission by emailing \href{mailto:info@vldb.org}{info@vldb.org}. Copyright is held by the owner/author(s). Publication rights licensed to the VLDB Endowment. \\
\raggedright Proceedings of the VLDB Endowment, Vol. \vldbvolume, No. \vldbissue\ %
ISSN 2150-8097. \\
\href{https://doi.org/\vldbdoi}{doi:\vldbdoi} \\
}\addtocounter{footnote}{-1}\endgroup

\ifdefempty{\vldbavailabilityurl}{}{
\vspace{.3cm}
\begingroup\small\noindent\raggedright\textbf{PVLDB Artifact Availability:}\\
The source code, data, and/or other artifacts have been made available at \url{\vldbavailabilityurl}.
\endgroup
}

\section{Introduction}

When data scientists have multiple datasets, each contains various variables, and scientists want to know what they could get from the data, determining the outcome after joining multiple relational tables. Assume all the datasets augmenting are valid candidates, the values are integers, no overlapping values with different attributes and each has no missing number value; scientists often want to find the goal from the datasets after augmenting in which they want to know what data can do with or predicting what based on the datasets, for example housing prices are often based on location (which contains near urban center or a place where the criminal status is low) and goverment policies (which has taxes incentive) or the home values ( including the age and size of the house). To predict the housing price, scientists with their expertise in the field, rank each variable in each dataset one at a time until there is no variable in any dataset remaining and augment the datasets which are location, government policies in the area, home values. 

The process of using machine learning application and causal relationship tasks to predict the outcome requires a utility function that users have to know exactly beforehand. This would be abundant for users to walk them through the process of finding utility functions based on data in datasets when users do not know their exact information. We solve this problem by introducing BOD: Blindly Optimal Data Discovery by eliminating the finding utility function process and by asking a small number of questions based on variables of datasets to output a small subset of rightful tuples, which is the goal of predicting the future. 

\textbf{The work of algorithms}: The query first asks the user to pick the most important variable in each dataset. For example, if a user want to predict housing price without knowing their utility function, the query asks user to rank each variable in each of the datasets once at a time in multiple rounds. Based on the user's ranking, the query selects all the tuples that meet the user's preferences based on utility score, and the process continues working with these selected tuples. The query then outputs all the desirable tuples whenever user stops ranking the variables. 

\begin{figure}
  \centering
  \includegraphics[width=\linewidth]{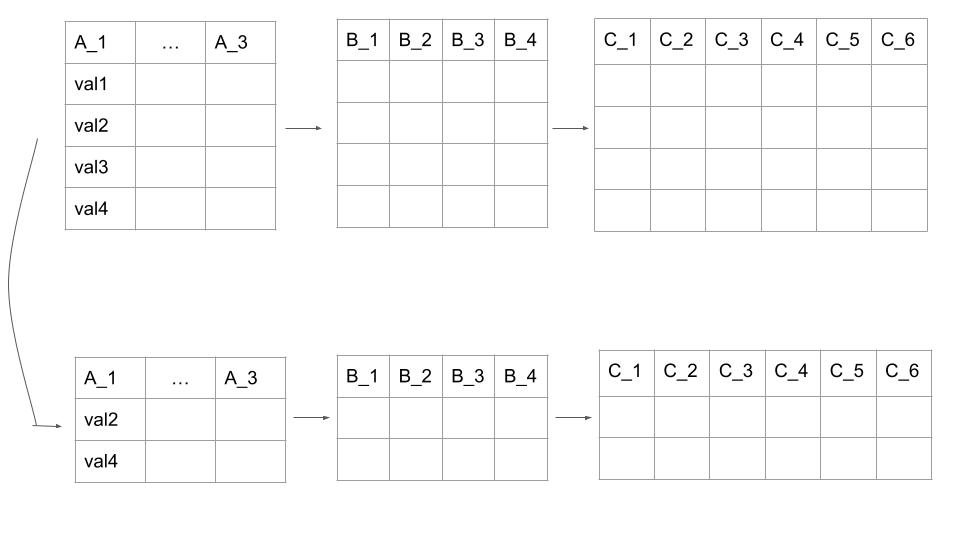}
  \caption{Augmenting data from tables and filtering out items based on human feedback of their ranking for variables}
\end{figure}

\section{Related Studies}

\textbf{Machine Learning.} To leverage data effectively for prediction or understanding what insights the data can offer, the technique of using machine learning as proposed in \cite{Galhotra2023MetamGD} requires the user's utility function. However, this approach is limited when the user's utility function is unknown. In such cases, the work of \cite{Gong2021VerVD} only selects a nearly optimal solution without fully understanding the potential of the data. Our algorithm, the Blindly Optimal Data Discovery, combines the advantages of these two approaches—enabling the discovery of data insights without requiring the user's utility function.

\textbf{Top-\textbf{k}.} The Top-K algorithm has lasted with many developments over time, as highlighted in \cite{FM12}, \cite{LYH09}, \cite{LC09}, \cite{QY12}, \cite{SI07}, \cite{Soliman2007TopkQP}, \cite{Khosla2015TopkQP}, and \cite{Xiao2016ProbabilisticTR}. Despite these advancements, the algorithm requires the user's utility function, which can be difficult to obtain and is not always applicable where user preferences are not well-defined.

\textbf{Skyline.} Numerous studies have explored this algorithm, including \cite{LYH09}, \cite{LY07}, \cite{MC09}, \cite{CJ06}, \cite{CJ062}, \cite{TX07}, \cite{TD09}, \cite{PT05}, \cite{GY05}, \cite{XZ08}, and \cite{Ying2020ANS}. A drawback of the Skyline algorithm is the uncontrolled size of the output, which can include uninteresting tuples. In contrast, it may exclude tuples dominated by others but still of interest to the user. To address this issue, researchers proposed combining the techniques of Top-\textbf{k} and Skyline, as seen in \cite{MLT21}, to control the output size more effectively.

\textbf{K-Dominance.} This algorithm, explored in \cite{Chan2006FindingKS}, \cite{Dong2014FindingKS}, and \cite{Siddique2012EfficientKS}, offers a method for refining the results by considering the dominance of data points across multiple dimensions, providing a balance between comprehensiveness and output size.

\textbf{Pareto Optimality.} Pareto optimality is central to multi-objective optimization, with contributions from studies such as \cite{Yu2019EfficientCC} and \cite{Geilen2005AnAO}. This approach seeks to identify optimal solutions that cannot be improved on one objective without worsening another, offering a rigorous framework for decision-making in complex scenarios.

\textbf{Blindly Optimal Data Discovery.} Our algorithm involves human interaction with the data to rank the variables within each dataset iteratively. After each ranking phase, we filter out promising tuples based on their utility scores, progressively refining the results for optimal data points without requiring predefined utility functions.

\section{Problem Definition}

\begin{table*}[t]
  \caption{Notation and meaning}
  \label{tab:commands}
  \begin{tabular}{|c|c|}
  \hline
  \centering
  $D$ & The set of all tuples\\
  $d$ & The total attributes of the final table after augmenting all the datasets\\
  $S$ & The table of all the relational tables augmented over $n_i$ attributes \\
  $t$ & The specific tuple that has an instance schema \\
  $y_{min_i}$ & The maximum values's summation based on user's $i$th ranking \\
  $y_{max_i}$ & The summation of all values based on user's $i$th ranking \\
  $y_i$ & The summation of all values of the instance\\
  \hline
  \end{tabular}
\end{table*}

Let $(S_1, \dots, S_n)$ denote a relation table over $n$ attributes, $S_i$ denotes the $i$th attribute. $D$ composes the list of tuples with a table $(S_1, \dots, S_n)$ such that every tuples $t \in T$ is an instance of the table.

\textbf{Linear function}: 

\begin{itemize}
    \item $LINEAR = \{f|f(x) =  \sum\limits_{i=1}^{d}f(x_i) \\ \text{\{where each} \ f(x_i)  \text{ is a linear function.}\}$

\end{itemize}

Assume we want to predict the housing price in an area, we have datasets including location, government policies in the area, and home values. In detail, the dataset location composes a relational table that has two attributes that are Near Urban and Criminal Free, the dataset government policies composes table that has one attribute which is Tax, while the dataset home values composes a tables that has two attributes which are Size and Age of the house. After augmenting all the datasets, we have a dataset composes five attributes and six instances after augmenting the tables as shown in the table 2 

\begin{center}
\begin{table}[h!]
\begin{tabular}{|l|c|c|c|c|c|}

\hline
Price & \multicolumn{1}{l|}{Near Urban}& \multicolumn{1}{l|}{Criminal Free} & \multicolumn{1}{l|}{Tax}& \multicolumn{1}{l|}{Size}& \multicolumn{1}{l|}{Age}  \\ \hline
House 1  & 26 & 5& 90  & 1500 & 150\\ \hline
house 2 & 35 & 2& 20 &  1300 & 120\\ \hline
House 3 & 45 & 3& 78& 2000 & 95\\ \hline
House 4 & 9 & 2& 46 & 1700 & 50\\ \hline
House 5 & 6 & 4& 65 & 1800 & 25\\ \hline
house 6 & 47 &7 & 30 & 1450 & 75\\ \hline

\end{tabular}
\caption{Example of the tables after augmenting}
\label{table:}
\end{table}
\end{center}

Then we scale down the values in the range of $\{0,1\}$ by the highest value in each variables and we calculate $y_{min_i}$ after asking the user to rank the variables in the original tables as shown in table 3.

\begin{center}
\begin{table}[h!]
\begin{tabular}{|l|c|c|c|c|c|c|c|}

\hline
Price & \multicolumn{1}{l|}{Near Urban}& \multicolumn{1}{l|}{Criminal Free} & \multicolumn{1}{l|}{Tax}& \multicolumn{1}{l|}{Size}& \multicolumn{1}{l|}{Age}& \multicolumn{1}{l|}{Sum} \\ \hline
House 1  & 0.55 &0.71& 1  & 0.75 & 1 & 4.01\\ \hline
house 2 & 0.74 &0.29 & 0.22 &  0.65& 0.8 & 2.7\\ \hline
House 3 & 0.96 & 0.43& 0.87& 1 & 0.63 & 3.89\\ \hline
House 4 & 0.19 & 0.29& 0.51 & 0.85 & 0.33 & 2.17\\ \hline
House 5 & 0.13 & 0.57& 0.72 & 0.9 & 0.17& 2.49 \\ \hline
house 6 & 1 & 1& 0.33 & 0.73 & 0.5 & 3.56\\ \hline

\end{tabular}
\caption{Example of the query scaling down all the values in all variables after augmenting }
\label{table:}
\end{table}
\end{center}

Suppose the user rank Near urban , Tax, and Size as the first priority in the first interaction with the algorithm, we then calculate the summation of each instance and find the maximum calculation in this round, which is $y_{min_1}=2.83$ in this case as shown in table 4.

\begin{center}
\begin{table}[h!]
\begin{tabular}{|l|c|c|c|c|}

\hline
Price & \multicolumn{1}{l|}{Near Urban} & \multicolumn{1}{l|}{Tax}& \multicolumn{1}{l|}{Size} & \multicolumn{1}{l|}{Near Urban + Tax + Size} \\ \hline
House 1  & 0.55  & 1  & 0.75 & 2.3\\ \hline
house 2 & 0.74 & 0.22 &  0.65& 1.61 \\ \hline
House 3 & 0.96 & 0.87& 1 & 2.83\\ \hline
House 4 & 0.19 & 0.51 & 0.85 & 1.55\\ \hline
House 5 & 0.13 & 0.72 & 0.9 & 1.75\\ \hline
house 6 & 1 & 0.33 & 0.73 & 2.06\\ \hline

\end{tabular}
\caption{Example of the query when user rank Near Urban, Tax, and Size as the first priorities }
\label{table:}
\end{table}
\end{center}

Based on the information from the first interaction, we find the tuples that have $y_{min_1} \leq y_1 \leq y_{max_1}$ that are House 3 and House 6 since $2.83 \leq y_1 \leq 3.89$. The table 5 shown the result from the first interaction. The final interaction which is the second ranking would result in only House 3.

\begin{center}
\begin{table}[h!]
\begin{tabular}{|l|c|c|c|c|c|c|c|}

\hline
Price & \multicolumn{1}{l|}{Near Urban}& \multicolumn{1}{l|}{Criminal Free} & \multicolumn{1}{l|}{Tax}& \multicolumn{1}{l|}{Size}& \multicolumn{1}{l|}{Age}& \multicolumn{1}{l|}{Sum} \\ \hline
House 3 & 0.96 & 0.43& 0.87& 1 & 0.63 & 3.89\\ \hline
house 6 & 1 & 1& 0.33 & 0.73 & 0.5 & 3.56\\ \hline

\end{tabular}
\caption{The results of the algorithm}
\label{table:}
\end{table}
\end{center}

\section{BOD: Blindly Optimal Data Discovery}

Since we scale down each value in a variable by a factor of max value of the variable such that every value now in the variable is in the range of $\{0,1\}$ and we scale the values in every variable after augmenting all the datasets. For each of the relational tables, We ask the user to pick the first most important variable in each of dataset, knowing that the number of variables in each table is varied, called $n_i$. We then find the maximum sum of values from the most first important variables from the original tables after augmenting, called $y_{min_1}$. Then, we find $y_{max_1}$ from the row with the maximum sum $y_{min_1}$, which is the sum of all values in that row. After that, we select the items that have their utility score, $y_1$ , which is the sum of all variables, in which $y_{min_1} \leq y_1 \leq y_{max_1}$. We then use these items in a new dataset and begin the second phase, and so on, until we get to the last variable in any table or the user wants to stop ranking the variables in each dataset. 

\begin{algorithm}
 \caption{Blindly Optimal Data Discovery}
 \label{Blindly Optimal Data Discovery}
 \begin{algorithmic}[1]
 \REQUIRE The tables $S_i$ that has $D$ tuples.
 \ENSURE return a subset $t$ of $D$.
 \STATE Let $S_i=1$ 
 \WHILE{$S_i \leq$ length of $S$}
  \FOR{$i$ in the length of $S_i$}
  \STATE Let the user rank the attributes in each table $S_i$ until the user stops 
  \ENDFOR
  \STATE  $S_i+=1$
 \ENDWHILE
 \STATE Find the maximum length of each $S_i$ in $S$, called $max_{n_i}$
  \STATE Concatenate DataFrames/Tables horizontally along columns
  \STATE $d_i=1$
  \FOR{$d_i \leq d$}
  \STATE find max value called $S_max_i$
  \STATE Scale down all values by dividing $S_{max_i}$
  \STATE $d_i+=1$
  \ENDFOR
  \STATE Update the new $S$
  \FOR{$i$ in range $max_{n_i}$}
  \STATE Based one user ranking in $i$th time, calculate $y_{min_i}$ and $y_{max_i}$ accordingly 
  \STATE If {$y_{min_i} \leq y_i \leq y_{max_i}$}
  \STATE Select all tuples $t$ that satisfies and let update $S$
  \ENDFOR
\RETURN $t$
\end{algorithmic}
\end{algorithm}

\section{BOD's Performance Guarantee}

We show that Blindly Optimal Data Discovery can identify the optimal solution in exactly $d$ queries and returns a nearly optimal solutions in a maximum of $d$ queries since the number of questions depends on the amount of the total attributes after augmenting all the datasets. The performance runs far better than any previously known algorithms.

\section{Experimental Evaluation}

We assume the augmentation part is correct, with no error from the scientist’s domain; all values are valid, not missing, and all are in positive integers. We tested the algorithm using Google Collaboration, which runs on GL65 Leopard 10SCXK, an x64-based PC, on Microsoft Windows 11 Home Single Language. We test on three synthetic datasets, each dataset has 10000 tuples and a range of 3 to 9 variables. The values in each cell are randomly generated in the range of {1, 1000}. Then, We test on three synthetic datasets in the range of 5000, 10000, and 15000 tuples, each with six variables. The values in each cell are randomly generated in the range of $\{1,1000\}$. 

\begin{figure}
  \centering
  \includegraphics[width=\linewidth]{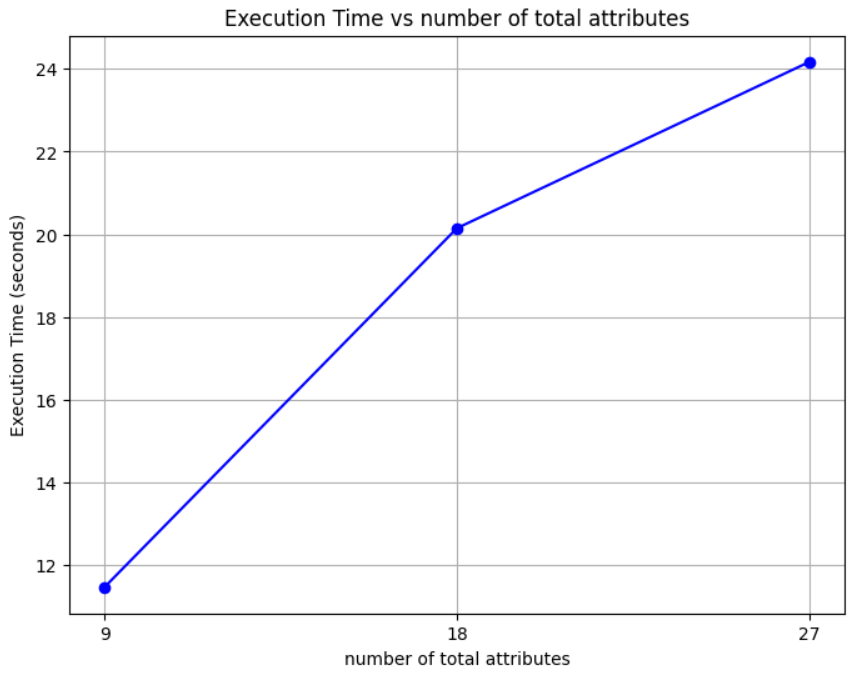}
  \caption{Changes in number of attributes}
\end{figure}

\begin{figure}
  \centering
  \includegraphics[width=\linewidth]{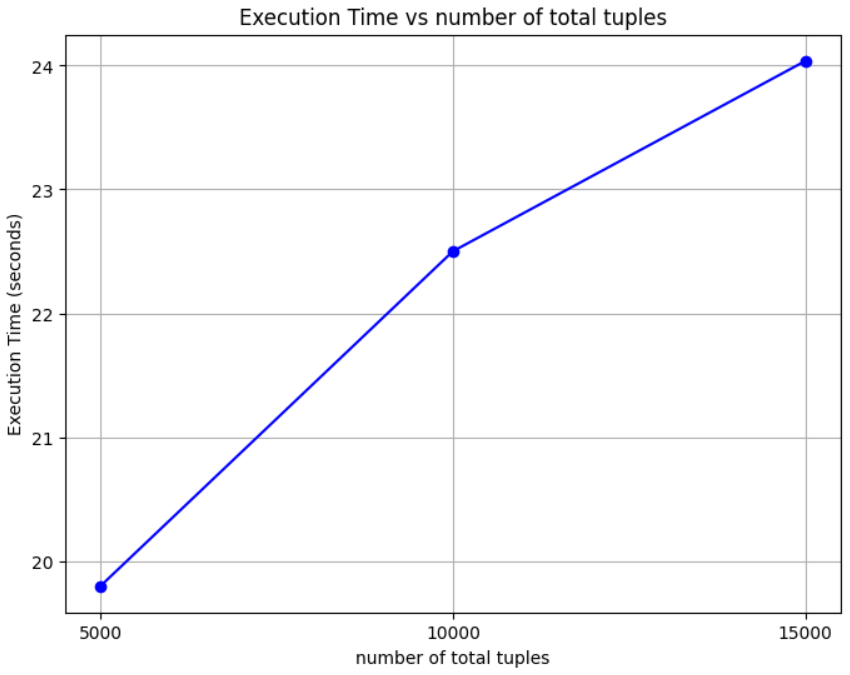}
  \caption{Changes in number of tuples}
\end{figure}

\begin{figure}[h]
    \centering
    \includegraphics[width=\linewidth]{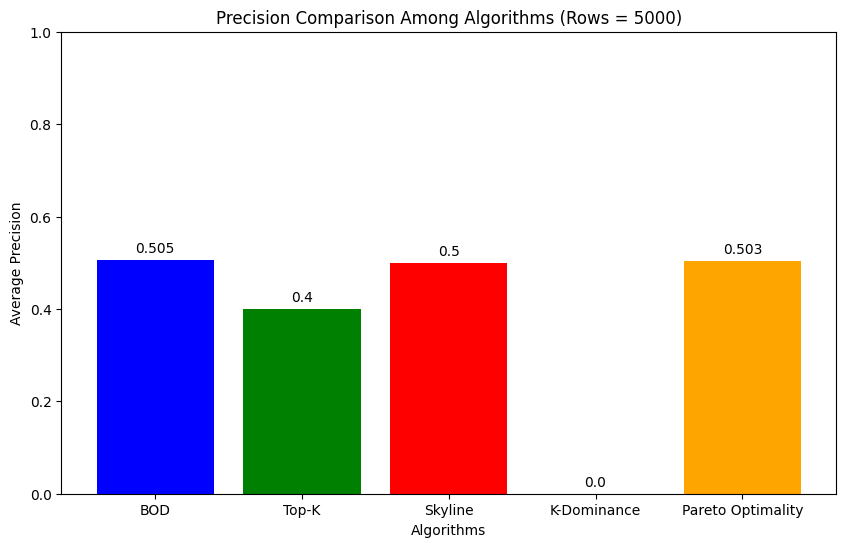}
    \caption{Precision Comparison Among Algorithms for 5000 Tuples}
    \label{fig:precision_compare_5000}
\end{figure}

The precision comparison among different algorithms, as depicted in Figure \ref{fig:precision_compare_5000}, provides important insights into the performance
of the Blindly Optimal Data Discovery (BOD) algorithm and
other established algorithms: Top-K, Skyline, K-Dominance, and
Pareto Optimality, when applied to a dataset testing on 5000 tuples. The BOD algorithm achieved a precision of 0.505, which outperforms the Top-K algorithm, which had a precision of 0.4. This result is important as it demonstrates that BOD can accurately identify relevant data points without requiring a predefined utility function, unlike Top-K, which is heavily dependent on the user’s utility function to rank the tuples. The slight advantage of BOD over Skyline (0.5) suggests that BOD not only matches but slightly exceeds the ability of Skyline to find optimal solutions under these conditions. The Pareto Optimality algorithm also performed closely to BOD, with a precision of 0.503. This similarity in performance highlights that both BOD and Pareto Optimality are capable of yielding compatible outcomes in terms of precision, even though they approach the
problem from different theoretical foundations. Pareto Optimality, which is observed in multi-objective optimization, and BOD, without predefined utility functions, with its application in ranking and filtering, achieve a high level of precision. However, the K-Dominance algorithm yielded a precision of 0.0 in this scenario. This let us know that K-Dominance may struggle with identifying relevant data points in larger datasets, particularly when the data is highly multidimensional or when the criteria for dominance are not easily satisfied. The results highlight the accuracy of the BOD algorithm in achieving high precision without requiring predefined user preferences, making it suitable in data discovery tasks. Its performance, on par with established methods like Pareto Optimality and surpassing Top-K and Skyline, under certain conditions, underscores its potential as a reliable alternative for scenarios where utility functions are unknown or difficult to define.

\section{Analysis of MAE Comparison Among Algorithms}

\begin{figure}[h]
    \centering
    \includegraphics[width=\linewidth]{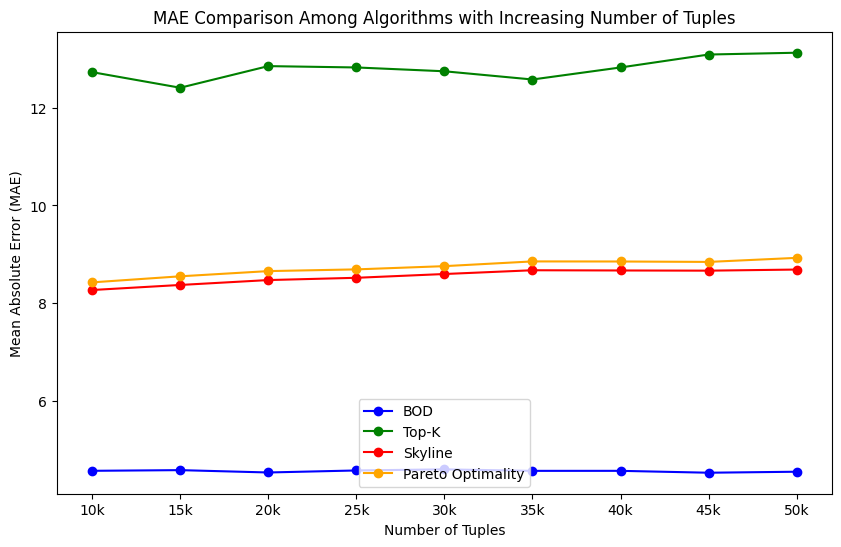}
    \caption{ Mean Absolute Error (MAE) Comparison Among Algorithms with increasing number of Tuples}
    \label{fig:mae_comparison_ChangesTuples}
\end{figure}

The results, as depicted in Figure~\ref{fig:mae_comparison_ChangesTuples}, let us know about that the BOD algorithm consistently achieved the lowest MAE across all dataset sizes which demonstrates that BOD is the most effective in selecting values that are closest to the actual housing prices.  On the other hand, the Top-K algorithm shows the highest MAE score among the four algorithms tested which is a deviation from the actual housing prices.  The Skyline and Pareto Optimality algorithms demonstrated intermediate performance, with MAE values that were consistently higher than those of BOD but lower than those of Top-K. Thus, BOD has the most precise accuracy among all algorithms.

\section{Analysis of Algorithm Precision and Stability}

\subsection{Mean Precision Comparison}
\begin{figure}[H]
    \centering
    \includegraphics[width=\linewidth]{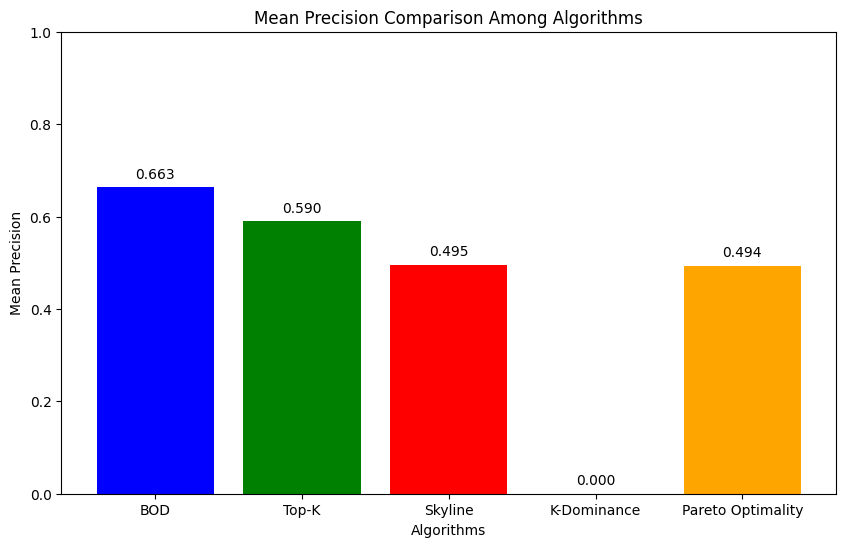}
    \caption{Mean Precision Comparison Among Algorithms}
    \label{fig:mean_precision}
\end{figure}

As shown in the chart in Figure \ref{fig:mean_precision}, among the algorithms, the Blindly Optimal Data Discovery (BOD) algorithm illustrates the highest mean precision of 0.663, which is the most effective at identifying relevant data points under the given experimental conditions. The Top-K algorithm followed with a mean precision of 0.590, suggesting that it also performs reasonably well, although it requires knowledge of the user's utility function. The Skyline and Pareto Optimality algorithms yielded nearly identical mean precisions of 0.495 and 0.494, respectively, which means that both algorithms are somewhat less effective in this context as effectively as BOD or Top-K. The K-Dominance algorithm, however, failed to identify any relevant data points, resulting in a mean precision of 0.000.  Thus, BOD again is the most effective among all algorithms.

\subsection{Precision Stability Analysis}
\begin{figure}[H]
    \centering
    \includegraphics[width=\linewidth]{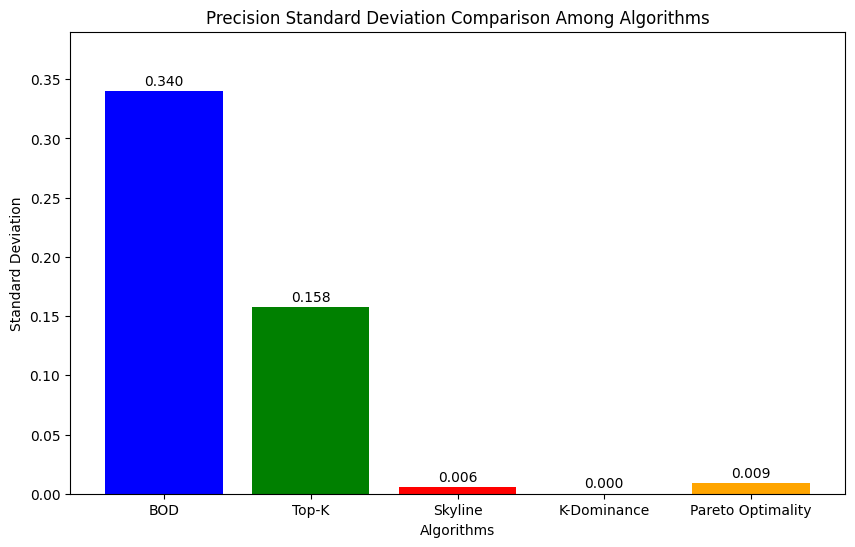}
    \caption{Precision Standard Deviation Comparison Among Algorithms}
    \label{fig:std_dev}
\end{figure}

The precision stability of each algorithm is presented in Figure \ref{fig:std_dev} across multiple iterations. Despite its high mean precision, the BOD algorithm shows a standard deviation of 0.340, which is somewhat variable and could be due to the sensitivity of the BOD algorithm to the specific characteristics of the dataset. In addition, While Top-K's mean precision was lower than BOD's, it demonstrated better stability with a standard deviation of 0.158. Skyline and Pareto Optimality algorithms illustrate very low standard deviations of 0.006 and 0.009, respectively. 

The analysis highlights a trade-off between precision and stability across the algorithms tested. BOD offers the highest precision but suffers from instability, while Skyline and Pareto Optimality provide highly stable but less precise outcomes. Top-K strikes a balance, offering moderate precision with relatively good stability. The K-Dominance algorithm's poor performance suggests it may not be suitable for the types of datasets and criteria used in these experiments.

\subsection{Testing on Boston Housing dataset}
\begin{figure}[H]
    \centering
    \includegraphics[width=\linewidth]{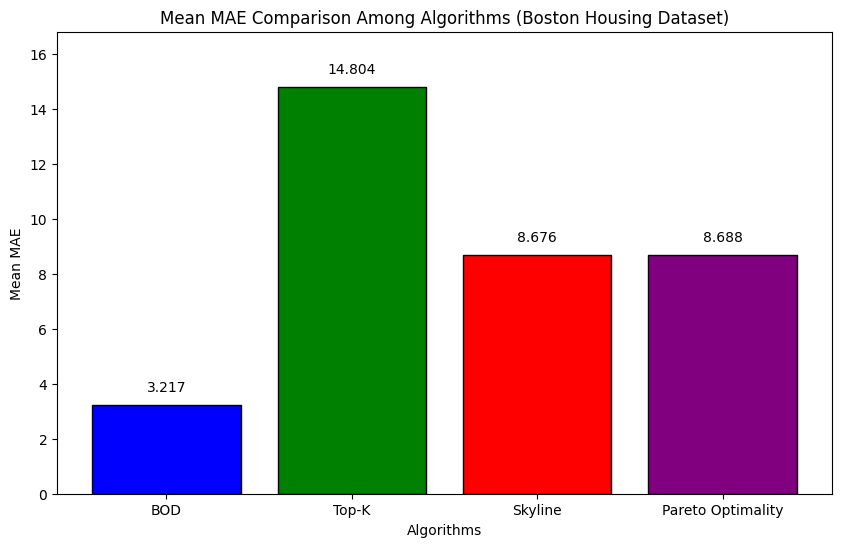}
    \caption{Mean MAE Comparison Among Algorithms (Boston Housing Dataset)}
    \label{fig:mae_comparison}
\end{figure}

The chart in Figure~\ref{fig:mae_comparison} \cite{nair2021boston} presents a comparison of the Mean Absolute Error (MAE) among four algorithms—BOD, Top-K, Skyline, and Pareto Optimality—applied to the Boston Housing Dataset. In detail, the BOD algorithm demonstrated superior performance, achieving the lowest mean MAE of 3.217, which is the most accurate prediction when compared to the true housing prices. In contrast, the Top-K algorithm showed the poorest performance with a mean MAE of 14.804, indicating significantly less accurate predictions. Both the Skyline and Pareto Optimality algorithms performed moderately, with mean MAEs of 8.676 and 8.688, respectively. The close MAE values for Skyline and Pareto Optimality suggest that these two algorithms provide similar levels of prediction accuracy, though they are less precise than BOD.

\section{Conclusion}

The BOD algorithm shows better performance compared to other algorithms such as Top-K, Skyline, and Pareto Optimality in terms of accuracy for the task of predicting housing prices with various datasets and environments. In addition, the BOD algorithm achieves lower Mean Absolute Error (MAE) and higher precision, which makes it an optimal choice for tasks involving complex decision-making and prediction. This also makes BOD valuable for applications such as housing price predictions and similar data-driven tasks for data scientists and analysts. Would open questions be better techniques/algorithms than BOD in terms of accuracy, scalability, and running time?

\section{Citations}

\bibliographystyle{ACM-Reference-Format}

\end{document}